ARTICLE   OPEN

# A comparative study using state-of-the-art electronic structure theories on solid hydrogen phases under high pressures

Ke Liao[1]*, Xin-Zheng Li[2,3], Ali Alavi[1,4] and Andreas Grüneis[5]*

Identifying the atomic structure and properties of solid hydrogen under high pressures is a long-standing problem of high-pressure physics with far-reaching significance in planetary and materials science. Determining the pressure-temperature phase diagram of hydrogen is challenging for experiment and theory due to the extreme conditions and the required accuracy in the quantum mechanical treatment of the constituent electrons and nuclei, respectively. Here, we demonstrate explicitly that coupled cluster theory can serve as a computationally efficient theoretical tool to predict solid hydrogen phases with high accuracy. We present a first principles study of solid hydrogen phases at pressures ranging from 100 to 450 GPa. The computed static lattice enthalpies are compared to state-of-the-art diffusion Monte Carlo results and density functional theory calculations. Our coupled cluster theory results for the most stable phases including C2/c-24 and P2$_1$/c-24 are in good agreement with those obtained using diffusion Monte Carlo, with the exception of Cmca-4, which is predicted to be significantly less stable. We discuss the scope of the employed methods and how they can contribute as efficient and complementary theoretical tools to solve the long-standing puzzle of understanding solid hydrogen phases at high pressures.

## INTRODUCTION

Hydrogen is the lightest and most abundant element in the Universe, yet its phase diagram at high pressures and low temperatures remains elusive. Due to the subtle interplay of quantum nuclear and electronic correlation effects,[1–6] the question as to which state of matter is stable at high pressures is controversial. Likely, candidates for high-pressure phases include various orientationally ordered molecular crystals,[7–13] (liquid) metallic,[14–22] superconducting[23] and superfluid systems.[24] These potentially exotic states of matter and their crucial importance for astrophysical, planetary as well as materials sciences has led to intensified investigations using both experimental and theoretical techniques. However, currently available calculated as well as measured equilibrium phase boundaries vary strongly with respect to the employed methods and suffer partly from uncontrolled sources of error.

Experiments that seek to determine properties of hydrogen under high pressures are hindered by various problems; for example, the low X-ray scattering cross section of hydrogen, the small sample sizes and the diffusive nature of hydrogen. Recent claims of experimentally measured metallic phases[22,25] are therefore under debate,[26] whilst earlier experimental results[15,27,28] have not been able to conclusively detect metallic behaviour up to a pressure of 320–342 GPa.

Determining the Wigner–Huntington transition[14] using theoretical methods is extremely challenging. Despite the significant advancements of modern ab initio theories in the past decades, the predicted metallisation pressure varies significantly in a range of ~150–450 GPa, depending on the employed method.[4,17,29–32] Most ab initio studies of solid hydrogen are based either on density functional theory (DFT)[12,17,29,33] or quantum Monte Carlo calculations.[2–4,31,32,34,35] DFT is considered the workhorse method in computational materials science, and can be used to calculate lattice enthalpies on the level of various approximate exchange and correlation (XC) energy functionals. Furthermore, the Hellmann–Feynman theorem provides access to atomic forces and allows for optimising structures, as well as calculating phonons on the level of DFT.[12] Calculated and measured infrared and Raman spectra serve as a reliable tool for a direct comparison between theory and experiment.[7–9,13,36–39] However, different parameterisations of the XC functional in DFT give inconsistent predictions, e.g., PBE predicts a too low metallisation pressure compared with experiments, whilst other exchange functionals produce higher pressures than DMC.[4,35,40]

Instead, more accurate methods including diffusion Monte Carlo (DMC) have been employed to predict more reliable pressure temperature-phase diagrams,[2–4,31,32] which correct the underestimation of the metallisation pressure by DFT-PBE to a large extent. However, DMC calculations rely on the fixed-node approximation, and most of the current studies use crystal structures optimised using DFT. A critical assessment of the errors introduced by these approximations is still missing in literature and requires computationally efficient and concomitantly accurate methods.

In this work, we show that quantum chemical wavefunction theories hold the promise to serve as an efficient and accurate tool for the investigation of high-pressure phases of solid hydrogen. In particular, we find that coupled cluster theory[41,42] achieves a good trade-off between computational cost and accuracy when employing recently developed techniques that allow for simulating the thermodynamic limit of periodic systems in an efficient manner.[43,44] We note that these finite size corrections have paved the way for a number of ab initio studies, including predictions of molecule–surface interactions[44–47] and pressure–temperature phase diagrams of carbon and boron nitride allotropes.[48] The studies referred to above have demonstrated that coupled cluster methods achieve a similar level of accuracy as DMC for solid state systems that are not strongly correlated. Moreover, coupled cluster methods have been benchmarked against various more accurate methods in model hydrogen systems,[49] showing the high accuracy of the methods in weakly correlated situations. Furthermore, we employ full configuration interaction Quantum Monte Carlo (FCIQMC)[50–52] in this work for small systems to examine the validity of the coupled cluster method.

[1]Max Planck Institute for Solid State Research, Heisenbergstrasse 1, 70569 Stuttgart, Germany. [2]School of Physics, State Key Laboratory for Artificial Microstructure and Mesoscopic Physics, Peking University, 100871 Beijing, P.R. China. [3]Collaborative Innovation Centre of Quantum Matter, Peking University, 100871 Beijing, P.R. China. [4]Department of Chemistry, University of Cambridge, Lensfield Road, Cambridge CB2 1EW, UK. [5]Institute for Theoretical Physics, Vienna University of Technology, Wiedner Hauptstrasse 8-10, 1040 Vienna, Austria. *email: k.liao@fkf.mpg.de; andreas.grueneis@tuwien.ac.at

### RESULTS

We investigate theoretical results for the static lattice enthalpies of solid hydrogen phases computed on different levels of theory. The static lattice enthalpy is defined by

$$H = E + PV,  \quad (1)$$

where $P$ is the pressure estimated from the $E-V$ relation and $V$ corresponds to the volume per atom. $E$ refers to the total ground-state energy per atom obtained using DFT, HF or CC theory in the Born–Oppenheimer approximation. In passing, we note that the importance of quantum nuclear effects for transition pressures of solid hydrogen phases has been explored in refs [1–6] In this work, we will focus on the accuracy of the employed electronic structure theories only, disregarding such contributions. The coupled cluster singles and doubles (CCSD) energy is defined as the sum of the Hartree–Fock and the corresponding electronic correlation energy.[42] We discuss the convergence of the electronic exchange and correlation energy contributions with respect to the employed $k$-mesh used to sample the first Brillouin zone and the one-electron basis set, as well as additional computational details in the supplementary information. The pressure–volume relation of each phase, $P(V) = -\frac{dE}{dV}$, is obtained in the following manner. The total energy retrieved as a function of the volume per atom, $E(V)$, is fitted with a polynomial function of $V^{-1}$ in an optimal order that minimises the fitting residual and provides smooth curves. We find that a third-order polynomial fitting is adequate for all phases, except for phase $P2_1/c$-24 which is fitted using a fourth-order polynomial. A further increase in the fitting order can result in artificial wiggling behaviours of the $H(P)$ curves. The derivative with respect to the volume is readily obtained in an analytic manner using the fitted $E(V)$ function, yielding smooth $P(V)$ curves. We present all static lattice enthalpies relative to the $C2/c$-24 phase unless stated otherwise. In total, we study five solid hydrogen phases: Cmca-4 (Cmca-Low), Cmca-12, $C2/c$-24, $P2_1/c$-24 and $P6_3/m$-16, where we have adopted the convention of naming the structures by their symmetries followed by the number of atoms in the primitive cells. Phase Cmca-4, Cmca-12 and $C2/c$-24 consist of layered hydrogen molecules whose bonds lie within the plane of the layer, forming distorted hexagonal shapes. Whereas some bonds of hydrogen molecules in phase $P6_3/m$-16 lie perpendicularly to the plane of the layer. $P2_1/c$-24 consists of molecules arranged on a distorted hexagonal close-packed lattice.

These structures have previously been selected as potential candidates as the most stable high-pressure phases of hydrogen[12] and have been studied by DMC methods. We notice that a family of 'mixed' structures are also identified as promising candidates in ref.,[12] however, for the current comparative studies among CCSD, DFT-PBE and DMC, they are not included here but could be an interesting topic for future work.

We have optimised the geometries of the structures employing the DFT-PBE functional.[53] The DFT calculations have been performed using the Vienna ab initio simulation package (VASP) employing a plane wave basis set in the framework of the projector augmented wave method.[54] More details about the structures can be found in refs. [2,12]

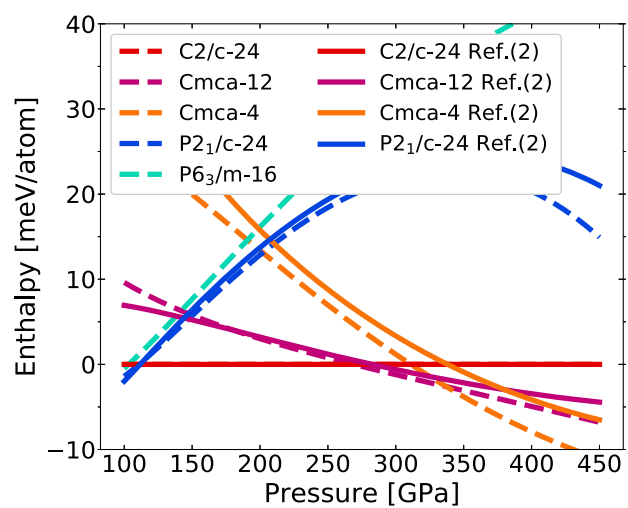

**Fig. 1** DFT-PBE relative enthalpies. The DFT-PBE relative enthalpies of structures that are used for CCSD calculations in this paper (dashed lines) and that of the structures from ref. [2] (full lines). DFT favours the atomic phase Cmca-4 at high pressures

We first discuss results of the investigated high-pressure phases on the level of DFT. Figure 1 depicts the DFT-PBE static lattice enthalpies relative to the $C2/c$-24 phase. DFT-PBE predicts the $C2/c$-24 phase as the most stable phase at pressures ranging from ~100–290 GPa. In a small range of pressures ~300 GPa, the Cmca-12 phase is found to be thermodynamically stable, whereas the metallic Cmca-4 phase becomes stable at pressures exceeding ~330 GPa. Experimentally, no metallic phases have been observed in this pressure range, and quantum nuclear effects do not account for this discrepancy either.[4] The too low metallisation pressure can be attributed to the lacking of van der Waals interactions in PBE functional,[40] resulting in underestimation of the stability in the molecular structures. We note that Fig. 1 also depicts static lattice enthalpies from ref. [2] obtained using DFT-PBE. We attribute the minor differences between the static lattice enthalpies to small differences in the employed structures and the fitting procedure that is employed to compute the lattice enthalpies from the total energies retrieved as a function of the volume per atom. We stress that the computed enthalpies are very sensitive to the employed structures.

In contrast to approximate XC functionals employed in DFT calculations, quantum chemical many-electron methods allow for approximating the electronic XC energy in a more systematic manner, albeit at significantly larger computational cost. The simplest wavefunction-based method is the Hartree–Fock (HF) approximation that neglects electronic correlation effects by definition, employing a single Slater determinant as Ansatz for the electronic wavefunction. Figure 2 depicts the static lattice enthalpies computed in the HF approximation relative to $C2/c$-24. In contrast to DFT-PBE results, we find that HF theory significantly reduces the stability of the Cmca-4 and Cmca-12 phases, shifting their transition pressures far above 400 GPa. However, the HF method is not a good approximation for metallic systems, despite the fact that it is free from self-interaction errors. In particular, HF band gaps are usually significantly overestimated compared with the experiment. Moreover, the lack of electronic correlation in the HF Ansatz leads to the neglect of van der Waals contributions that are crucial for a correct description of relative stabilities of molecular crystals.[55] We note that van der Waals contributions to the binding energy of molecular crystals become in general larger for smaller volumes due to the polynomial decay of the dispersion interaction with respect to the intermolecular distance. Due to the reasons outlined above, the static lattice enthalpies calculated on the level of HF theory are expected to exhibit significant errors compared with more accurate electronic structure theories and will serve as a reference for post-HF methods only.

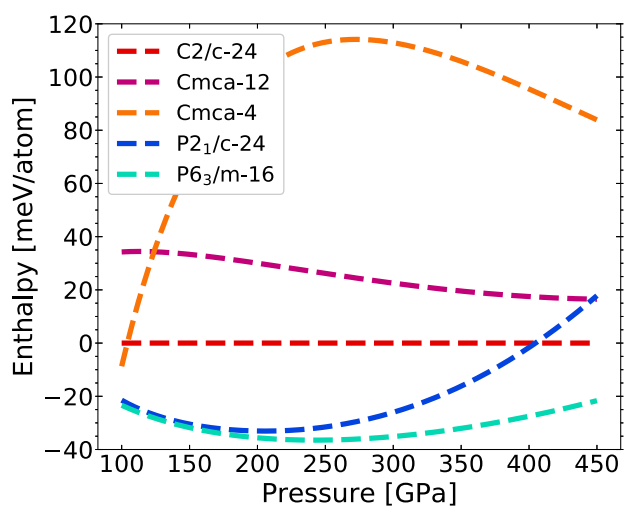

**Fig. 2** HF relative enthalpies. The HF relative enthalpies of structures that are used in this paper. In contrast to the DFT-PBE result, the atomic phase Cmca-4 is unfavoured at high pressures

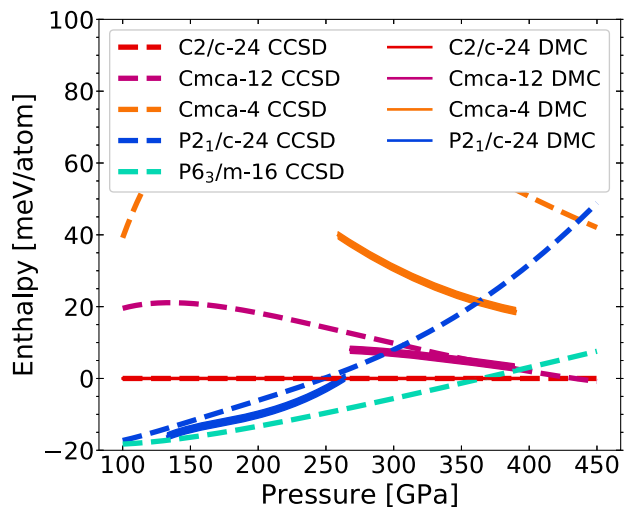

**Fig. 3** CCSD relative enthalpies. The CCSD relative enthalpies of structures that are used in this paper (dashed lines) and the DMC relative enthalpies of structures from ref. [2] (full lines). The thickness of the full lines refer to the standard deviations of stochastic sampling of the 1st Brillouin zone while performing twist-averaging in the DMC calculations. In this work, the 1st Brillouin zone is sampled using a dense regular grid such that the errors are converged to within 1 meV/atom. See supplementary information for details. CCSD and DMC[2] agree very well in the most stable molecular phases, i.e., C2/c-24, P2$_1$/c-24 and Cmca-12, whilst the only discrepancy exists in the Cmca-4 phase, which is predicted by DFT-PBE to be metallic at high pressures. The phase transition between P6$_3$/m-16 and C2/c-24 predicted by CCSD happens at ~350 GPa, which agrees reasonably well with the DMC transition pressure range 250–350 GPa from ref. [3]

Here, we employ the coupled cluster singles and doubles (CCSD) method to account for electronic correlation effects using a HF reference. Periodic CCSD theory results for the static lattice enthalpies relative to the C2/c-24 phase are shown in Fig. 3. Compared with the HF theory, CCSD stabilises the Cmca-4 phase by ~40 meV/atom at pressures above 300 GPa. Similarly, the relative static lattice enthalpy of Cmca-12 is lowered by ~20 meV/ atom in CCSD compared with HF. For the P2$_1$/c-24 and P6$_3$/m-16 phases, we observe an opposite effect of the CCSD correlation energy contribution, reducing their stability relative to C2/c-24 by ~30 meV/atom at pressures exceeding 250 GPa. We note that CCSD theory reduces the differences in the relative static lattice enthalpies of the considered phases compared with the HF approximation.

The CCSD energy is the sum of the HF energy and an approximation to the electronic correlation energy that is computed using an exponential Ansatz for the wavefunction. Due to the many-electron nature of the employed Ansatz, CCSD theory is exact for two-electron systems. The coupling between electron pairs is, however, approximated by truncating the many-body perturbation expansion in a computationally efficient manner and performing a resummation to infinite order of certain contributions only.[42] As a consequence, CCSD theory is expected to yield highly accurate results for the molecular hydrogen crystals. This is confirmed by comparing with the corresponding DMC results from ref. [2] for C2/c-24 and P2$_1$/c-24 depicted in Fig. 3 that agree very well with our CCSD findings. Furthermore, static lattice enthalpies obtained on the level of CCSD and DMC (only shown in ref. [2]) for Cmca-12 relative to C2/c-24 are in good agreement as well and the transition pressure between P6$_3$/m-16 and C2/c-24 by CCSD (≈ 350 GPa) and DMC (≈250–350 GPa only shown in ref. [3]) are in reasonable agreement. However, we note that the DMC and CCSD results differ by ~40 meV/atom for the relative static lattice enthalpy of the Cmca-4 phase. In particular, the difference of the static lattice enthalpies of Cmca-4 and C2/c-24 at 350 GPa are ~100 meV/atom, 60 meV/atom and 20 meV/ atom using HF, CCSD and DMC, respectively.

## DISCUSSION

We now discuss possible reasons for the discrepancy between DMC and CCSD results for the Cmca-4 phase. DMC calculations employ the fixed-node approximation, whereas CCSD theory truncates the particle–hole excitation operator in the exponent of the wavefunction Ansatz. Fixed-node DMC gives the upper bounds[56] to the total energies of each phase. However, it is not necessarily the case that the lower enthalpy difference between Cmca-4 and C2/c-24 predicted by DMC is more reliable than that by CCSD, since the fixed-node errors in each phase do not necessarily cancel out accurately. The fixed-node errors in the total DMC energy can be estimated using backflow transformations and by comparing with full configuration interaction quantum Monte Carlo (FCIQMC)[50,52] results for the uniform electron gas.[57,58] It has been shown that the fixed-node errors are ~1 mHa per electron (27.2 meV/electron) in the high-density regime. In the case of solid hydrogen, the authors of ref. [35] report in their Supplementary Material that the energy in phase C2/c-24 is lowered by 1 mHa/atom (27.2 meV/atom) when employing back-flow transformations and ref. [3] reports that for Cmca-4 the backflow transformations lower the energy by only 10 meV/atom. This indicates that backflow transformations can depend significantly on the phases. Even though a large part of the fixed-node errors are expected to cancel when the energy difference between phases is computed, the remaining errors can still be on the scale of 10 meV/atom. On the other hand, we stress that the change from HF to CCSD relative static lattice enthalpies is on the scale of 40 meV/atom, indicating that a better approximation to the many-electron wavefunction than employed by CCSD theory could be necessary to achieve the required level of accuracy. We have also performed calculations using higher level theories, including FCIQMC, for smaller supercells containing 24 atoms at volumes corresponding to a DFT pressure of 400 GPa. These findings indicate that post-CCSD corrections to static lattice enthalpy differences for Cmca-4 and P2$_1$/c-24 are expected to be ~10 meV/atom. In short, both DMC and CCSD rely on good cancellations in errors introduced by their respective approximations to produce accurate predictions, especially when phases of different physical natures are compared. In addition to the inherent errors of DMC and CCSD theory, finite size and basis set errors can also be significant. The latter only applies to CCSD calculations and has been checked carefully as outlined in the supplementary information. As regards the finite size error, we study supercells containing 96 atoms and employ twist averaging as well as structure factor interpolation methods for our CCSD calculations to achieve a level of precision that is comparable with DMC results. Despite the above considerations, we can currently not draw any firm conclusion about the reason for the discrepancy between DMC and CCSD results for Cmca-4. However, we note that recently developed basis set convergence acceleration techniques will enable future studies of bigger systems using CCSD[59] and FCIQMC[60] theory that can hopefully provide more insight.

Despite the discrepancy between CCSD and DMC findings for Cmca-4, we point out that the good agreement for the static lattice enthalpies of the most stable high-pressure hydrogen phases is encouraging. Achieving accurate thermodynamic limit results for such systems on the level of CCSD theory has only become possible recently due to the development of the corresponding finite size corrections as outlined in refs [43,44] Furthermore, we note that the computational cost of the corresponding CCSD calculations is still moderate compared with methods with a similar accuracy. A single CCSD ground-state energy calculation for a system containing 96 atoms using 400 bands requires ~250 CPU hours, implying that it will become possible in the near future to perform structural relaxation of the employed crystal structures rather than relying on structures optimised using DFT-PBE. This is necessary for truly reliable predictions of high-pressure phases of solid hydrogen.

We have presented static lattice enthalpies for high-pressure phases of solid hydrogen calculated using state-of-the-art electronic structure methods, including coupled cluster theory. We find that CCSD theory results agree well with DMC findings from ref.:[2] phase C2/c-24 becomes more stable than phase P2$_1$/c-24 at ~250 GPa; phase Cmca-4 and Cmca-12 are less stable than phase C2/c-24 in the pressure range from 100 GPa to 400 GPa. The only discrepancy between CCSD and DMC is found for the Cmca-4 phase, and we have discussed possible sources of error. Future work will include the effects of the nuclei motions which are crucial in making theoretical predictions comparable with experiments. Based on the presented findings, the required computational cost of the employed CCSD implementation and recent methodological advancements,[59] we conclude that prospective CCSD studies will make it possible to optimise structures of solid hydrogen phases at high pressures with DMC accuracy. This will enable complementary CCSD and DMC studies with a significantly improved level of accuracy and achieve unprecedented physical insight into the Wigner–Huntington transition of solid hydrogen.

## METHODS
The CCSD calculations have been performed employing the coupled cluster for solids (CC4S) code interfaced to the Vienna ab initio simulation package (VASP). The projector augmented wave method, as implemented in VASP,[54,61,62] is used for all calculations. This section provides an overview of the computational methods and convergence techniques employed in this work. For more details, we refer the reader to the corresponding sections in the supplementary information under the same section titles.

### Geometries
The structures have been optimised using DFT-PBE and are similar to those employed in ref.[12] The forces on the atoms of the optimised structures are not larger than 0.1 eV/Å. With hindsight, it would have been preferable to use exactly the same structures as published in ref.[2] However, for the purpose of this work, the agreement between the structures suffices. For the CCSD calculations, we employ supercells containing up to 96 atoms that are as isotropic as possible and are obtained using the same method as described in the supplementary Note 2 of ref.[2] In this manner, finite size errors can be significantly reduced.

### CCSD basis set convergence
For the equilibrium phase boundaries in the pressure–temperature phase diagram, only relative enthalpies are relevant. Therefore, we have converged the energy differences with respect to the basis set only. MP2 natural orbitals (MP2NOs)[63] provide faster convergence than canonical Hartree–Fock orbitals (HFOs) computed from the full plane wave basis set. The convergence tests of the CCSD correlation energy differences with respect to the number of orbitals per atom relative to phase C2/c-24 have been carried out using supercells containing 24 atoms for all phases, except for phase P6$_3$/m-16 which contains 16 atoms in the supercells. The results are summarised in Table I and Fig. 1 in the Supplementary Information File. We note that the basis set incompleteness errors are mainly due to the electronic cusp conditions, which are very local effects and are not dependent on the supercell size.[64]

### Hartree–Fock finite size convergence
The HF energies are converged to within 1 meV/atom using increasingly large supercells or dense $k$-meshes sampling the first Brillouin zone. The required system sizes for all phases are summarised in Table II in the Supplementary Information.

### CCSD finite size convergence
The twist-averaging technique[65] and finite size corrections,[43,44] based on the interpolation of the transition structure factor, are applied on 96-atom supercells to approximate the thermodynamic limit of the CCSD correlation energies.

### Post-CCSD error estimates
We applied some higher level theories, including DCSD,[66,67] CCSD(T)[42,68] and FCIQMC, to estimate the post-CCSD error. The results are summarised in Fig. 2, Fig. 3, Fig. 4 and Table III in the Supplementary Information File.

## DATA AVAILABILITY
The data that support the findings of this study are available from the corresponding authors upon reasonable request.

## CODE AVAILABILITY
The VASP code is a copyrighted software and can be obtained from its official website. The CC4S code is available from A.G. upon reasonable request and will be made open-source in the future.

## ACKNOWLEDGEMENTS

We thank Pablo López Ríos for providing the code to generate isotropic supercells, the original DMC static enthalpy-pressure data and some useful discussions. The computational results presented have been achieved in part using the Vienna

Scientific Cluster (VSC) and the IBM iDataPlex HPC system HYDRA of the Max Planck Computing and Data Facility (MPCDF). Support and funding from the European Research Council (ERC) under the European Unions Horizon 2020 research and innovation program (Grant Agreement No. 715594) is gratefully acknowledged.

## AUTHOR CONTRIBUTIONS

A.G. designed and led the research; K.L. performed the research; A.G. and K.L. wrote the paper; X.Z.L. provided the crystal structures and A.A. advised and provided the tools to perform the FCIQMC calculations.

## COMPETING INTERESTS

The authors declare no competing interests.

## ADDITIONAL INFORMATION

**Supplementary Information** is available for this paper at https://doi.org/10.1038/s41524-019-0243-7.

**Correspondence** and requests for materials should be addressed to K.L. or A.G.

**Reprints and permission information** is available at http://www.nature.com/reprints

**Publisher's note** Springer Nature remains neutral with regard to jurisdictional claims in published maps and institutional affiliations.